# Moisture-induced Damage Evolution in Laminated Beech

Mohammad Masoud Hassani, Falk K. Wittel, Samuel Ammann, Peter Niemz, Hans J. Herrmann

Institute for Building Materials, Department of Civil, Environmental and Geomatic Engineering, ETH Zurich, Stefano-Franscini Platz 3; CH-8093 Zurich
Corresponding author: (Falk K. Wittel) fwittel@ethz.ch - Phone: +41 44 633 28 71

## Abstract
Structural elements made of laminated hardwood are increasingly used in timber engineering. In this combined numerical and experimental approach, damage onset and propagation in uni-directional and cross-laminated samples out of European beech due to climatic changes are studied. The inter- and intra-laminar damage evolution is characterized for various configurations adhesively bonded by three structural adhesive systems. Typical situations are simulated by means of a comprehensive moisture-dependent non-linear rheological finite element model for wood with the capability to capture delaminations. The simulations give insight into the role of different strain components such as visco-elastic, mechano-sorptive, plastic, and hygro-elastic deformations under changing moisture content in progressive damage and delamination. We show the stress buildup under cyclic hygric loading resulting in hygro-fatigue and modify an analytical micro-mechanics of damage model, originally developed for cross-ply laminates, to describe the problem of moisture-induced damage in beech lamellae.

## Introduction
New innovative timber engineering applications harness the improved mechanical performance of hardwood. High stiffness and strength of beech for example make it possible to achieve comparably slender, wide spanning structures with respect to softwood. However, the disadvantageous low dimensional stability and inhomogeneity of beech make it necessary to adhesively bond smaller sized pieces to form large laminated structural members. After several years of intense research on adhesive bonding of hardwood, it becomes evident that the traditional framework, manifested in codes for adhesive bonding of softwood, is hardly applicable to hardwood. Typically, for beam or plate structures the longitudinal axes (L) of lamellae are oriented coaxially or perpendicularly to each other. Climatic changes or moisture gradients result in high residual stresses that often exceed values of the yield stress or strength of the bulk wood and adhesive bonding alike. In real structures differential moisture contents of components prior to gluing can significantly contribute to the development of residual stresses. Consequently, various types of cracks form which are defined by the normal orientation of the crack plane and the corresponding propagation direction. The weakest crack systems in wood are commonly TL, RL and TR thus, crack planes with the normal directions in tangential (T) and radial (R) direction (first letter) that propagate along the grain (L) or in the radial direction (second letter) (Wittel et al. 2005; Silva et al. 2006; Fortino et al. 2012; Qiu et al. 2014).

Adhesive bond-lines can be considered as internal interfaces with distinct properties from adherends and adhesives and strong gradients in material behavior (Hass et al. 2012). Cracks can penetrate adhesive bond-lines or can be deflected into it what results in delaminations. This is determined by mechanical performance of the bond-line due to adhesive, cohesive, or interphase fracture (Serrano 2000; Simon and Valentin 2003; Fortino et al. 2012; Hass et al. 2013). The conditions that lead to damage and its propagation, however, cannot be understood if the available energy for crack formation, hence the stress state and elastic energy are not known. Unfortunately, experiments alone will not be able to provide this insight as the observed deformations are a combination of reversible hygro-elastic, irreversible plastic and history-dependent visco-elastic and mechano-sorptive strain components. To make matter worse, most constitutive parameters of wood and adhesives exhibit significant non-linear dependence on the moisture content rendering intuitive explanations questionable (Hering 2011). However, numerical simulations e.g., with Finite Element Methods (FEM), can cope with the arising complexity of the coupled hygro-mechanical problem. Since they are based on continuum assumptions, effects of disorder - present on all hierarchical scales of wood - are smeared out. One obtains so to say the average answer over many components or samples. Even though it might be off with respect to a very single specific sample, the much more important general behavior as well as physical insight are gained.

In this manuscript, we study the damage evolution in small, three-layered beech wood panels of various thicknesses, adhesively laminated in aligned or crosswise way by three structural adhesives with distinct mechanical behavior based on a combined experimental and numerical approach. Intra- and inter- laminar damage develop and evolve above a certain thickness as freely supported samples cycle through drying and moistening conditions. We make the analogy to the well-studied problem of progressive transverse ply cracking in cross-ply composite laminates (Wittel et al. 2003) and interpret our observations for cross-laminated wood elements in the framework of the energy based micro-mechanics of damage approach (Nairn 1989; Liu and Nairn 1990; Nairn and Hu 1992). Unfortunately, this approach is limited to linear hygro(thermo)-elastic, transversely isotropic material bodies. To capture the complicated moisture-dependent rheology of wood however, advanced material models are required, like the recently introduced model that combines moisture-mechanical simulations with a hygro-elastic material including multi-surface plasticity, linear visco-elasticity and mechano-sorption, all considering directional and moisture-dependent parameters (Hassani 2015). Only if one does not neglect the relevant rheonoumous and scleronomous strain portions that contribute to the total strain, local stresses and strain energy densities can be calculated correctly and failure criteria can be used to make correct predictions. With this approach, the important problem of hygro-fatigue that is puzzling timber engineers in the form of spontaneous structural failures, even decades after erection, naturally emerges by sequential stress buildup.

This paper is organized as follows: In the Materials and Methods section we give a detailed description on the experimental campaign followed by a brief introduction to the rheological models. Model parameters that were not fully described in Hassani (2015), however, are summarized in the Appendix. In the results section, we first discuss the experimental observations on various configurations after the first two moisture cycles. We characterize the damage evolution by typical properties common to intra- and inter-laminar fracture studies. We verify FEM simulations on the macroscopic behavior of a set of experiments, before we apply the model to gain a deeper view on the mechanical situation that leads to the observed damage evolution. Hence, we first artificially introduce cracks in the middle lamella and next calculate the change in the strain fields and stresses for the onset of delamination as well as its propagation using fracture mechanics. These simulations are very detailed with respect to the material behavior. To be able to make more general statements on the damage evolution and scaling, we adopt a variational mechanical approach to the moisture-induced damage evolution studied in this manuscript. As the experimental time frame is limited to 2 cycles only, we extend it by simulations to 10 cycles (3 years) to be able to quantify the stress buildup for longer moisture histories. Finally, we conclude on the applicability of this study to delamination testing of adhesive systems.

## 2. Materials and Methods

### 2.1. Experimental tests

Three-layered European beech samples of various configurations were produced to undergo cyclic climatic changes. All samples had dimensions of L = 100 mm x 100 mm, but different thickness of 2d with d=4, 10, 20, 30 mm being named T1 to T4, respectively. They were either glued with parallel grain (P1) or cross-wise (P2) with the L-direction of the middle lamella of thickness d being perpendicular to the ones of the outer layers with thickness d/2 each (see Fig. 1). For adhesive bonding, three commonly used adhesive systems in timber construction, namely melamine urea formaldehyde (MUF), phenol resorcinol formaldehyde (PRF), and one-component polyurethane (1C PUR) were applied. The wood was conditioned and glued at 20° C/95% relative humidity (RH), following the specifications given in Table 2. To avoid drying during cold curing in the press machine under 1.2 MPa, samples were sealed from the environment by wrapping them in foil. For each configuration T1-T4, P1-P2 and adhesive type MUF-PUR-PRF, 4 samples were produced, resulting in 96 samples for testing (see Table 1). Just to give an example, sample P2-PUR-T3-2 refers to the second cross-laminated sample with 20 mm thick middle lamella glued by PUR adhesive. Note that bonding with MUF and PRF at 20° C/95% RH can result in starved bond-lines, while PUR performs outstandingly when bonded under wet conditions. Nevertheless, in the current study this approach for the model experiments was chosen to obtain strong tensile stresses upon drying.

**Table 1** Test specimen catalogue

|  |  | T1 | T2 | T3 | T4 | Adhesive type |
|---|---|---|---|---|---|---|
|  | Thickness d (mm) | <u>4</u> | <u>10</u> | <u>20</u> | <u>30</u> | MUF / PRF / PUR |
| Uni-directional (P1) / Cross-laminated (P2) | Number of samples | 4 | 4 | 4 | 3 |  |

All samples were placed in internally ventilated climate boxes for drying and remoistening. A moisture cycle comprised the following steps: (a) de-moistening from a nearly fiber saturated condition (95% RH) to 2% RH; (b) re-moistening to 95% RH. Drying was established using a layer of Silica gel and re-moistening by vaporized distilled water. The RH was recorded by means of a RH recorder located in each box. When the mass change was below 0.1%/day, the samples were considered to be equilibrated with the climate inside the box. At the end of every drying stage, the moisture-induced dimensional changes and deformations were recorded using a dial gauge. The climatic changes resulted in a change of moisture content of approximately 20%. Due to differences in the shrinkage behavior of adjacent layers, micro-cracks and delaminations were expected to form (see Fig. 1). The initiation and evolution of the inter- and intra-laminar cracks were recorded by scanning each side of every laminated sample with a flatbed scanner for further evaluation.

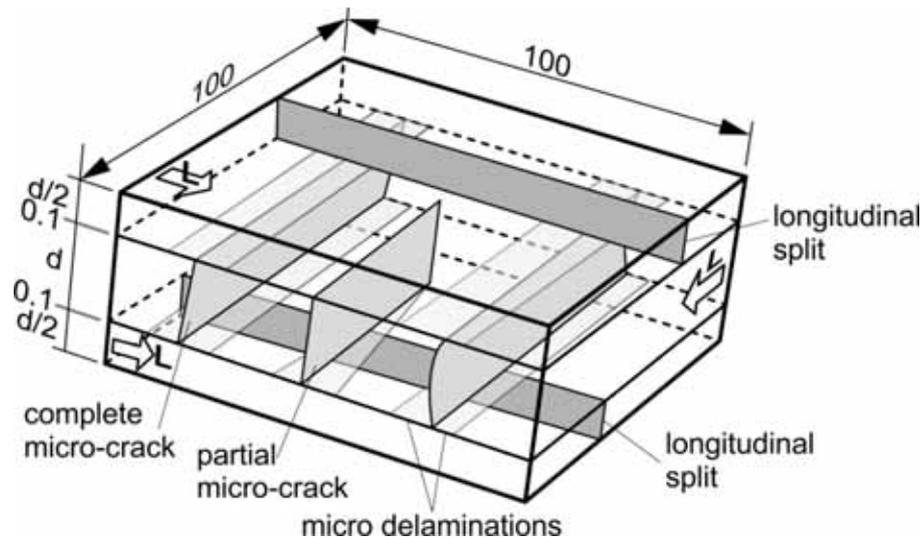

**Fig. 1** A cross laminated sample with dimensions in mm and different kinds of failure mechanisms (Nairn and Hu 1994)

## 2.2. Numerical procedure

For a better interpretation, we used a 3D orthotropic moisture-dependent constitutive material model for wood consisting of all instantaneous and history-dependent responses, namely hygro-elastic, plastic, visco-elastic, and mechano-sorptive deformation modes, previously developed and validated. For a full description of our implementation in the FEM package Abaqus we refer to Hassani (2015). The model gives for the first time the possibility to attain a better quantitative understanding of the true distribution and evolution of the moisture-induced stresses as the main driving forces for the formation of inter- and intra-laminar fracture. Note that the moisture analysis is only sequentially coupled to the non-linear mechanical analysis. The presence of adhesive bond-lines requires special treatment with respect to moisture transport and mechanical behavior.

In principle, a simplified moisture-dependent material model with respect to the one used for wood was employed for adhesives. In case of the PRF, it was a simple isotropic, hygro-elastic model, for the MUF viscoelasticity was added, and for the PUR additionally plasticity, represented by the simple J2-plasticity with isotropic hardening. A complete set of moisture-dependent adhesive properties for the MUF, PRF and PUR are summarized in Tables 3-7 in the Appendix. Hence, in combination with the full model description in Hassani (2015), all required parameters are defined.

The five-layered FEM model consists of three solid wood lamellae and two adhesive layers. Cohesive elements connect the adhesives with the middle layer to allow for the simulation of the onset and evolution of delamination within the framework of non-linear fracture mechanics (Alfano and Crisfield 2001; de Borst et al. 2004). The cohesive zone models were characterized in terms of the so-called traction-separation relationships describing the damage initiation and evolution laws. In the current study, the COH3D8 8-node 3D cohesive elements of ABAQUS (ABAQUS 2014) with an uncoupled traction-separation approach were used. Delaminations initiated due to a maximum nominal stress criterion followed by an energy-based damage evolution law with exponential softening. Mixed-mode cases were considered by a power law interaction criterion with the exponent 1.6 taken from Fortino et al. (2012). Table 8 in the Appendix summarizes all cohesive parameters utilized to characterize the fracture process zone. The adhesive bond-line was modeled by volume elements of thickness 0.1 mm, while we assigned a thickness of 0.001 mm to the cohesive layer. Abaqus by default uses the constitutive thickness equal to unity in the calculations. To consider the influence of the actual thickness, the normal and shear stiffness of the cohesive elements, representing the slope of the linear part of the traction-separation laws, were divided by the geometrical thickness (see Table 8). Note that only average values of the adhesive moisture-dependent Young's and shear moduli in a range from 0-5% moisture content were considered, since the dry state is the critical one with respect to the delamination growth. For wood layers with significant curvature of the growth rings (d>10 mm), the material orthotropy was assigned in a cylindrical coordinate system by estimating the location of the center (pith) from the local curvatures of each layer obtained from the side scans.

The only change in the boundary conditions of the simulations was the moisture content applied to all external faces of the model. The relation to the experimentally measured RH is given by the sorption isotherm curves for solid wood and all adhesive types (Wimmer et al. 2013). It is assumed that moisture transport inside the system can be described by Fick's law (Fortino et al. 2009; Gereke 2009). The obtained moisture field evolution is used in the subsequent stress analysis with the non-linear material models. One obtains full access now to all relevant strain components and resulting stresses. To allow for simple generation of each individual sample with its respective material coordinate systems, a parameterized script-based approach was chosen. A typical model is shown in Fig. 2.

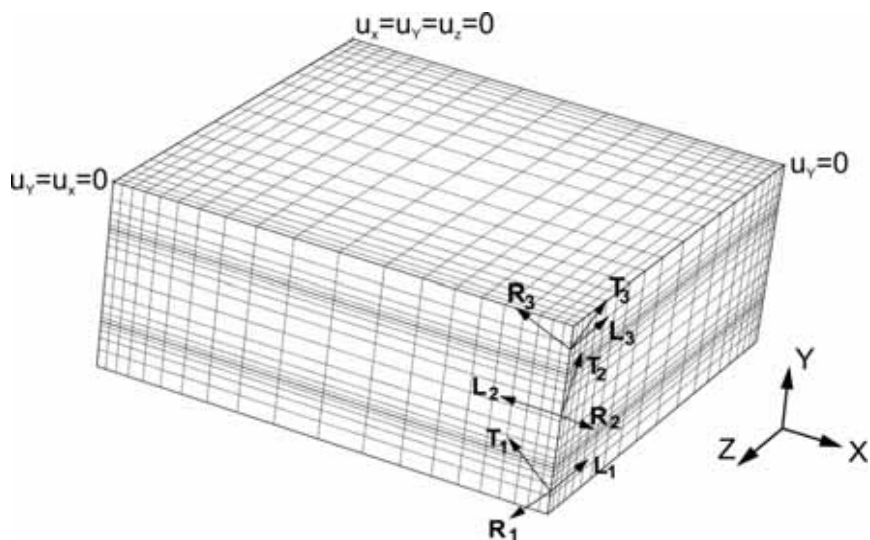

**Fig. 2** Exemplary FEM model with the applied boundary conditions and local material system consisting of twenty-node quadratic brick elements with reduced integration (C3D20R) and linear cohesive elements

## 3. Results

On a micro-mechanical scale, crack formation can be interpreted either as fiber cell wall failure or as fiber-debonding (Wittel et al. 2005). In principle, crack propagation by de-bonding consumes less energy, but is bound to fixed degradation planes, originating from the cellular arrangements. The large content of ray tissue of beech and the huge porosity due to the intra-ring vessel network (Hass et al. 2010) result in additional micro-mechanical damage mechanisms. In this study, however, cracks are addressed as if they would grow in a homogeneous, anisotropic medium. The expected types of intra-laminar failure are straight or curved RT or TL transverse micro-cracks that partially or entirely split the middle or outer lamellae. At the bond-line, cracks become inter-laminar by penetrating the bond-line or by being deflected along the interface resulting in delaminations, whatever mechanism is energetically favorable. A summary of the expected fracture types was added to Fig. 1.

## 3.1. Experimental observations

**Samples with coaxial L-directions (P1):** Lamellae with aligned grain orientations are typically found in structural glulam beams. Transverse cracks are usually not problematic, but micro delaminations from crack deflections into the interface significantly disturb the shear coupling of adjacent lamellae, weakening the entire element in terms of strength and stiffness. To reduce cracking, lamella thicknesses are limited by design codes and longitudinal stress relief cuts are often introduced.

Thin samples with a middle lamella thickness of d=4 mm (P1-MUF/PRF/PUR-T1) showed no sign of damage, but an excessive warping deformation that even increased in consecutive cycles. The dependence of the maximum warping deflection on the adhesive type and middle lamella thickness d is shown in Fig. 3. The moisture-induced deformations accumulate when cycling, resulting in hygro-fatigue. Also the accumulated magnitude of the moisture-induced warping depends on the mechanical performance of the adhesive type in order of MUF-PRF-PUR, what corresponds to the respective order of stiffness (see Table 3). The stiffest bond-line, therefore, has more hindering ability and lessens the deformation.

For all aligned samples with the middle lamella thickness d=10 mm (P1-MUF/PRF/PUR-T2), no damage was recorded at the end of the first drying step. P1-MUF/PUR-T2 samples remained flawless even after the second de-moistening while few TL-cracks appeared in the middle lamella of T2 samples laminated with PRF adhesive, pointing at the importance of the relative orientation of the longitudinally aligned lamellae material coordinate systems.

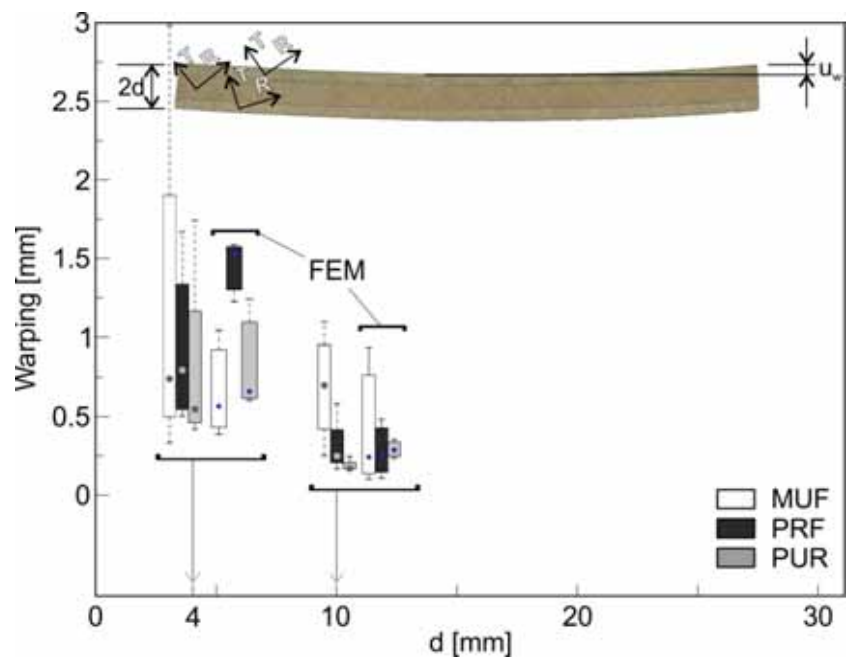

**Fig. 3** Dependence of the maximum warping deflection on the adhesive type and middle lamella thickness d averaged over 3 samples and side view of the deformed shape of the P1-PUR-T1-4 sample at the end of the second de-moistening step. In the following, the central box indicates the central 50% of a data set (limited by the lower and upper boundaries representing the 25% and 75% quantile of data). The central point shows the median and the dashed lines connect the two lower and upper markers that quantify the remaining data lying outside the central box.

For thicker samples with d=20 mm and d=30 mm (P1-T3/T4), mainly all middle lamellae cracked in the TL crack plane, primarily during the first de-moistening. In the second drying step those cracks mostly propagated towards the bond-lines. When reaching it, three different scenarios were observed: (a) crack arrest at the interface, (b) crack penetration through the bond-line and further propagation, (c) deflection into the interface (see Fig. 4). In principle, cracks propagate along the path that minimizes the energy (He 1994). Herein, the anatomic orientations of wood lamellae are a key issue, if the stress enhancement at the crack tip is not weakened by a very soft interface. From Fig. 4 we realize that when the weak TL planes of adjacent laminates are similar, cracks simply penetrate the interface, whereas for strongly different T-axes, it can be more favorable for cracks to be deflected and to grow along the bond-line, depending on the adhesive type. This results in micro delaminations or - if bond-lines are tough – in bond-line penetration, followed by crack growth along paths that are more energy consuming with respect to the growth in the fixed degradation plane of TL cracks. Note that also changes of the growth direction into the weak TL crack plane were observed at later stages.

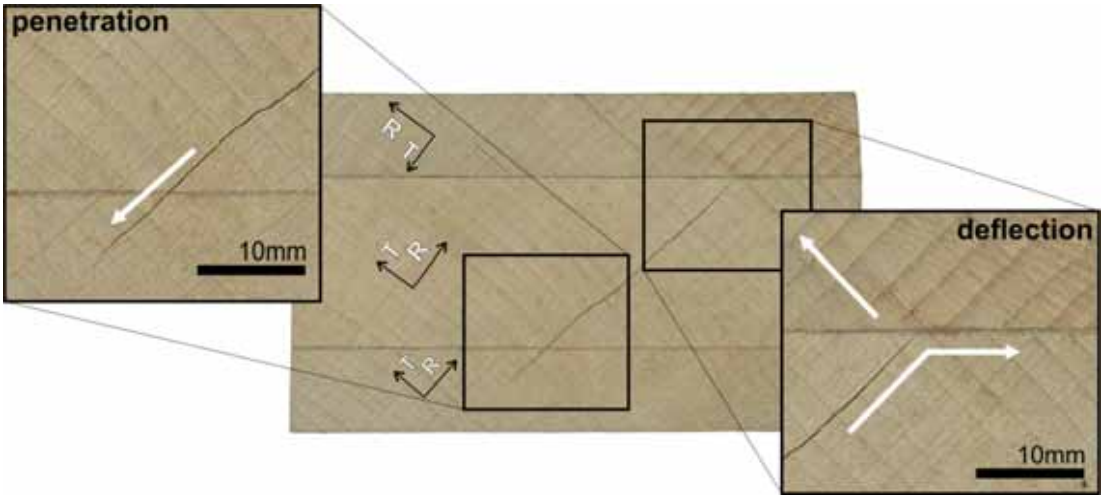

**Fig. 4** End grain surface views of the P1-PUR-T4-1 sample after the second drying, illustrating the deflected and penetrating cracks. Arrows mark the propagation trajectory after intersection with the interface

**Samples with crossed L-directions (P2):** Similar to the aligned ones, none of the T1 samples with d=4 mm showed any observable damage. As expected, the warping was significantly reduced in the crosswise configurations; however, the increased dimensional stability resulted in stronger edge face deformations (see Fig. 5) due to larger residual stresses. One can observe a similar dependence on the adhesive stiffness as for the warping deflection of the aligned samples.

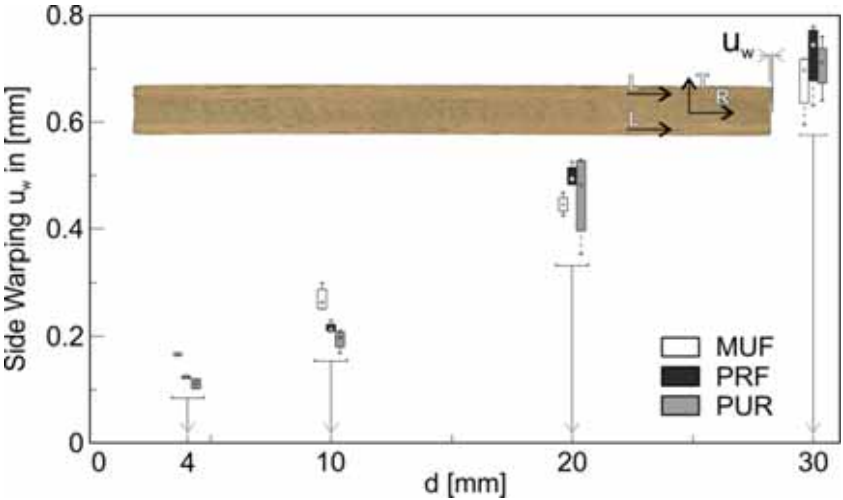

**Fig. 5** Side warping of all P2 samples after the first de-moistening step. Inset: sample P2-MUF-T1-2 at the end of the second de-moistening stage

With increasing thickness d, stresses due to drying exceeded strength values. For d=4 mm (T1) no damage was observed, while in d=10 mm (T2) samples, inter- and intra-laminar failure developed regardless of the material orientation and adhesive type. Fig. 6 shows the edge views of a sample (P2-MUF-T2-2) for two successive de-moistening stages. After the first de-moistening (left), an approximately evenly spaced set of inclined micro-cracks is observed. The quasi-periodicity is the result of a process where cracks form sequentially between existing cracks in accordance to transverse ply cracking in cross-ply laminates (Nairn 1989; Liu and Nairn 1990; Nairn and Hu 1992; Nairn and Hu 1994). As stresses in the middle layer are introduced by shear at the layer interfaces, they exhibit shear lag in a spatially limited zone. If the shear lag zone is smaller than the half crack spacing, an unperturbed zone exists where new cracks can initiate at the weakest location. If the shear lag zones interfere, cracks primarily form between existing cracks where the stress is maximal. In the second de-moistening, no further transverse cracks formed but microcrack-induced delaminations emanated from all micro-crack tips. Since no shear stress can be transmitted through the delaminations, the shear lag zones became closer and consequently the saturation crack density was reached (Nairn and Hu 1994).

The damage of the outer layers is best visible when the sample is rotated. It is characterized by staggered or anti-symmetric pattern of micro-cracks and adhesive failure (see Fig. 6). In the second drying phase, a new micro-crack initiated in a non-symmetric manner and additionally further de-bonding, originating from the micro-crack tips and edges formed. Samples with PUR and PRF adhesive behaved similar to MUF glued samples, but exhibited a smaller crack density in general. These observations are in agreement with typical crack formation processes in cross-ply laminates driven by a contrast in the Poisson's ratios and expansion coefficients of adjacent layers (Nairn and Hu 1992; Nairn and Hu 1994). Like for MUF samples, the crack opening increased in the second moisture cycle, pointing at the importance of the irreversible strains. It is interesting to note that cracks can be inclined since the formation of a longer crack along the weak TL plane can be energetically advantageous compared to a perpendicular crack that is shorter but requires more energy to grow.

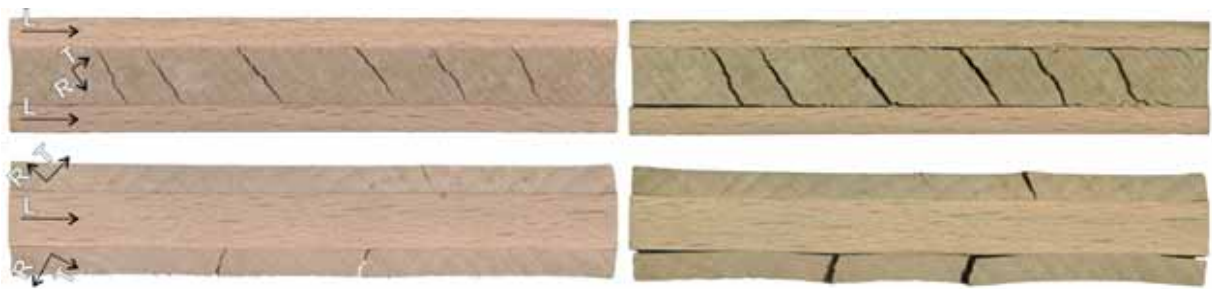

**Fig. 6** Side views of the fractured P2-MUF-T2-2 sample after the first (left) and the second (right) de-moistening steps, displaying the formation of equally distanced curved micro-cracks that span the entire sample width and microcrack-induced de-bonding

The general fracture behavior of P2-T3 samples (d=20 mm) was identical with their P2-T2 counterparts with the same adhesive system. The damage evolution is shown for the P2-PRF-T3-2 sample in Fig. 7. Note that regardless of the TL degradation plane, two more or less regularly spaced partially straight micro-cracks were formed.

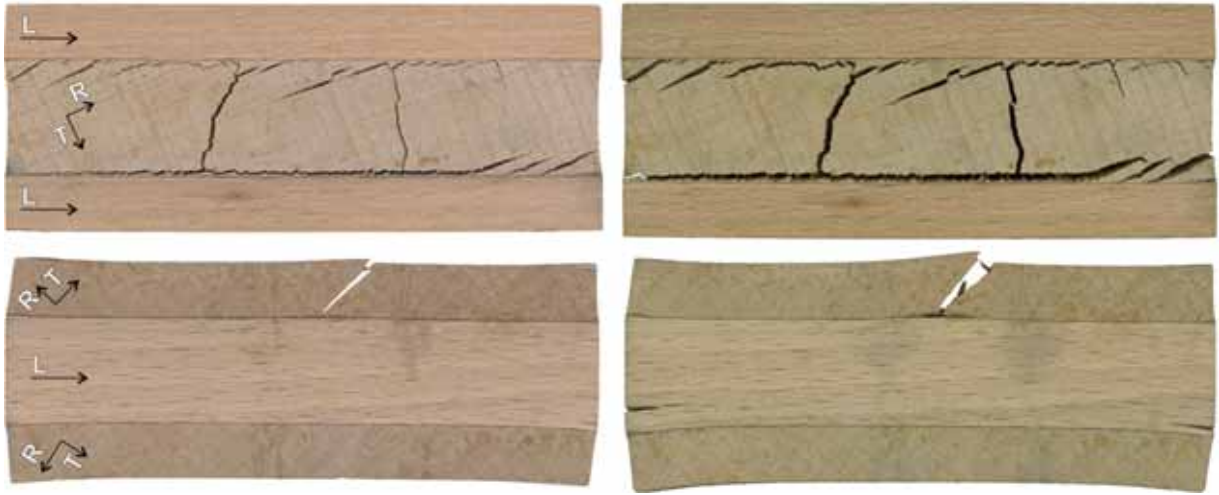

**Fig. 7** Side views of the fractured P2-PRF-T3-2 sample after the first (left) and the second (right) de-moistening steps with a pair of relatively straight transverse micro-cracks formed as well as a top layer crack with delamination

As the middle lamella thickness increased to d=30 mm (P2-T4) the analogy to transverse ply cracking breaks down, as the crack spacing reached sample dimensions. Eventually, growth in the weak TL-crack plane dominated, entirely splitting the middle lamella and thus, avoiding further transverse crack formation (see Fig. 8). The curvature of the growth rings, hence the location of the pith, plays a crucial role for intra- and inter-laminar damage as it dominates the overall stress state from the restrained shrinkage.

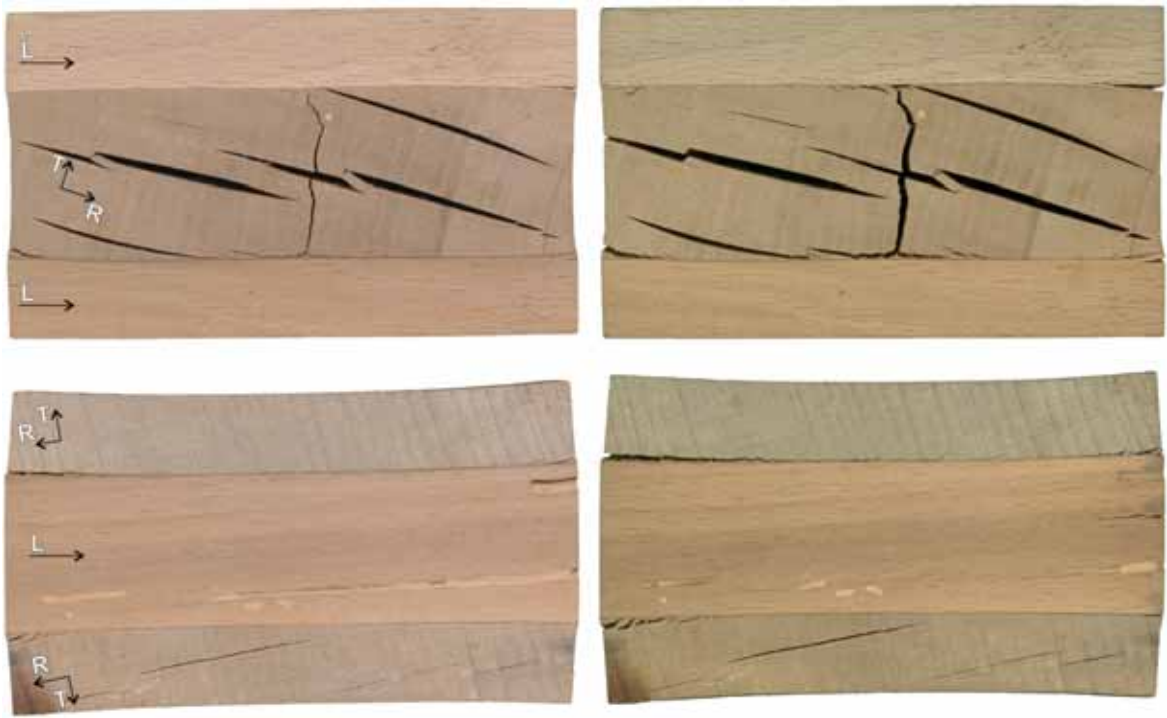

**Fig. 8** Side views of the fractured P2-PUR-T4-1 sample after the first (left) and the second (right) de-moistening steps with a single straight transverse micro-crack and severe intra-laminar TL-cracking

**Observations on bond-line performance:** The failure modes of the MUF and PRF/PUR bond-lines were characterized differently by adhesive failure and interphase delamination, respectively. MUF bond-lines predominantly exhibited adhesive failure for all thicknesses d. This is due to the high stiffness of the MUF resulting in a hard layer of the bulk adhesive incapable of reducing interfacial stresses. Moreover, the penetration of the MUF adhesive into the interphase layer increases the brittleness of the cellular structure and decreases the rigidity and the strength of the interface (Fortino et al. 2012). The superior performance of the PUR bond-line with fracture toughness higher than adjoining hard substrates can be associated with the occurrence of the plastic deformations within the adhesive material leading to a more flexible interface with the ability of reducing stress concentrations (River 2003). Note that differences also originate from a distinct liquid adhesive penetration behavior into the vessel network (Hass et al. 2012; Mendoza et al. 2012). Hence, under identical situations, the delamination is less pronounced (see Fig. 9). Similarly, the rigidity and the endurance of the PRF bonded-joints compared to the solid wood adherends can be attributed to the moderate stiffness and also the tendency of the PRF to receive more water resulting in a more plasticized and softer interface able to lower the stress concentrations all over the bond-line (River 2003). Note that our study shows similarities to delamination testing (DIN EN 14080) and analogously one could measure the wood fracture portion by injecting ink and tearing the sample mechanically apart.

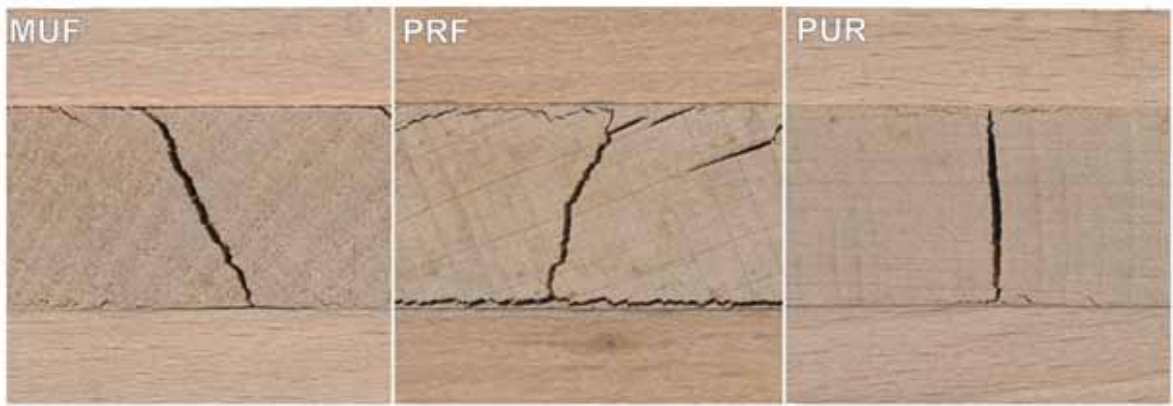

**Fig. 9** P2-X-T3 samples (d=40 mm) with different adhesives after the first de-moistening exhibiting adhesive failure in the MUF and interphase failure in the PRF and PUR bond-lines

The damage state is best quantified by the micro-crack density and the delamination ratio defined as $0.5(d_1+d_2) \cdot a^{-1}$ (see Fig. 10). One can observe the tendency of increasing delamination ratios but decreasing crack density with increasing thickness d.

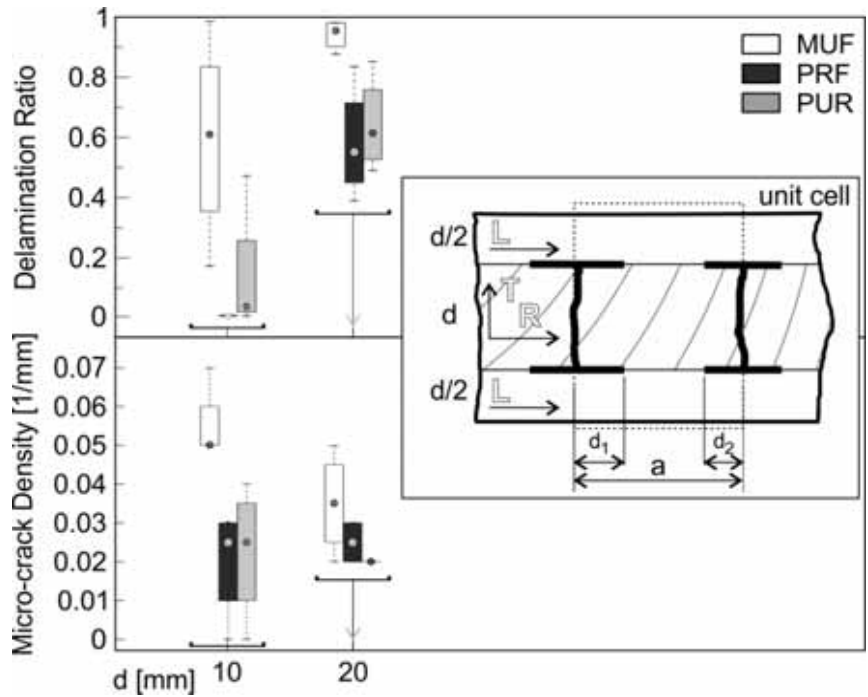

**Fig.** 10 Damage states after the second drying for different thicknesses and adhesive types. The inset gives the unit cell of damage used for the analytical stress calculation (see Sec. 3.3)

## 3.2. Numerical prediction of damage in cross-laminated samples under changing climate

As the moisture controls all hygro-mechanical properties, its gradients due to diffusive moisture transport as well as differential swelling are the main driving forces for the damage evolution in laminated wood. In a first step the moisture fields were calculated, exhibiting large gradients across the lamellae interfaces. The simulation was repeated for all adhesives using sorptive transport properties from Volkmer et al. (2012) and Wimmer et al. (2013) fitted with the equation and parameters summarized in Table 7. The resulting moisture profiles are basically identical, showing the irrelevance of the diffusive transport through adhesives compared to the one in the solid wood.

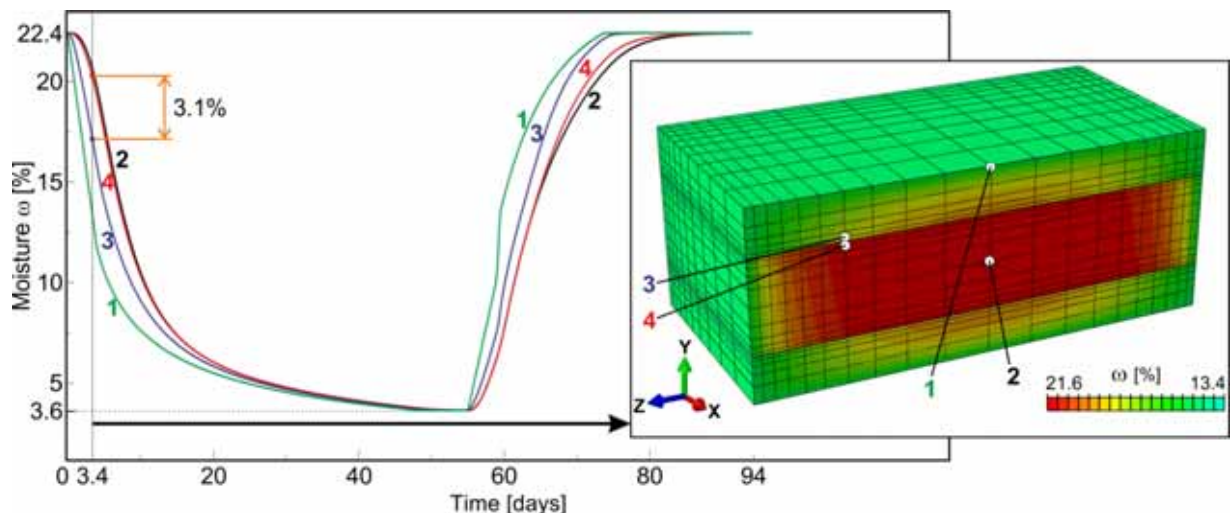

**Fig. 11** Moisture evolution at specific points and (inset) moisture field in the P2-T3-3 sample after 1.7 days

Already, we calculated the deflection with the material model using the previously determined evolution of the moisture field. This calculation was done for all coaxial T1/2 samples with their respective growth ring orientations. Good agreement is found with respect to the experimental values at the end of the second de-moistening step (see Fig. 3), even though material parameters are identical for all calculations what is clearly not the case for the experiments. Note that further verification examples for the material model can be found in Hassani (2015). In the following, the model is applied to predict the influence of the total strain constituents on the failure of cross-laminated configurations. To gain further insight into the damage onset and evolution, fracture mechanical simulations using cohesive interface elements are performed.

Delaminations initiate from existing transverse cracks, before they initiate at free edges (see Fig.14 right). This, however, does not mean that transverse cracks necessarily have to form before edge delaminations can initiate. To understand this, it is important to look at the situation for delamination initiation at a single crack. We simulate this case by artificially opening a crack at the dry state and then calculating a moistening and another de-moistening phase. It is striking to observe how the crack opening displacement (COD) of the transverse crack that forms in the first drying phase increases in the next step (see Fig. 12), driving micro delaminations. When the crack forms, the elastic strains are immediately relaxed and the relaxation of the visco-elastic strains starts. This is true for the cracked middle layer as well as for the compressively loaded outer plies. However, this does not explain the increased COD in the consecutive cycles. Provided no delaminations occur, the COD increases as visible in Fig. 12 by a decrease of the tensile mechano-sorptive strain during remoistening, resulting in an increased COD of about 25%. Additionally, as soon as the crack forms, the compressive plastic strains in the thickness directions develop (Fig. 12 top left in terms of the plastic hardening variable).

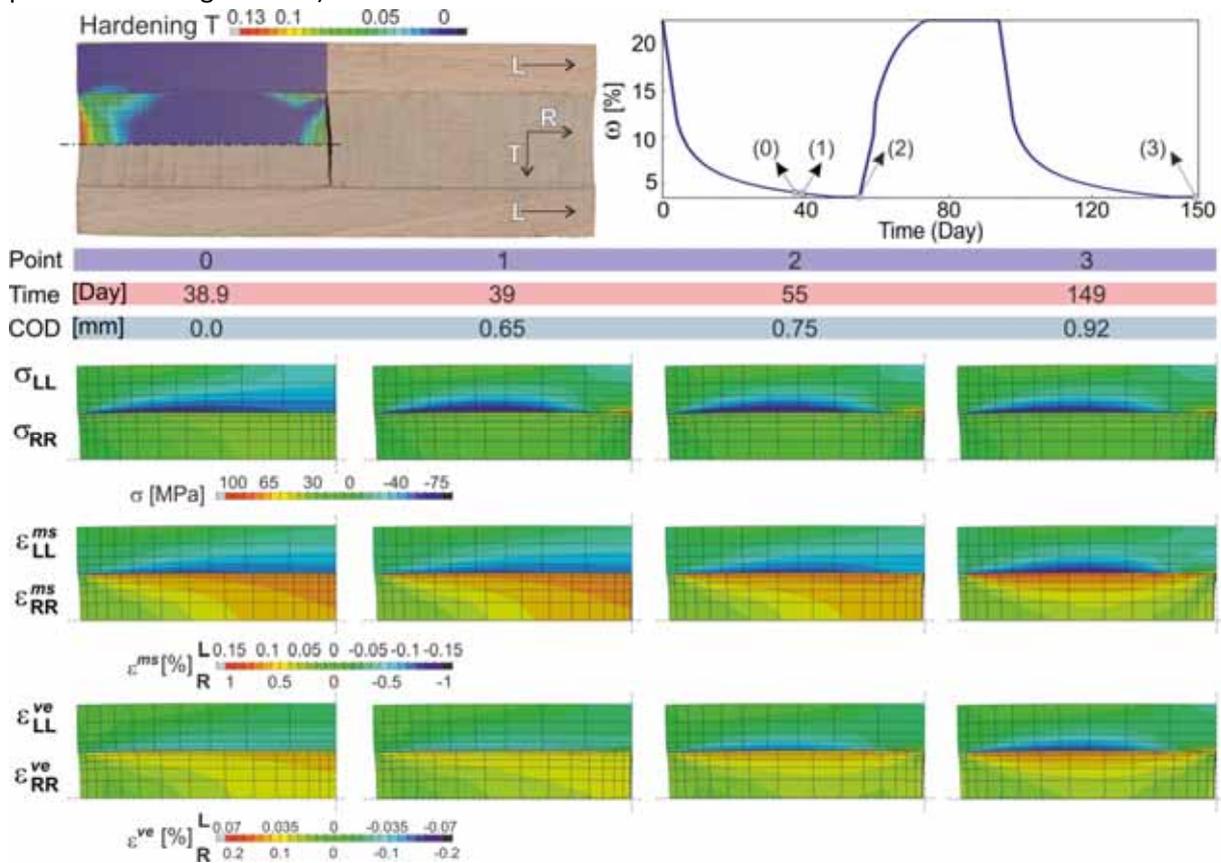

**Fig. 12** Strain evolution in quarter-models upon formation of a transverse crack with suppressed delamination for the P2-PUR-T3-3 sample with identical legends for the time evolution

Delaminations can initiate and grow under different fracture modes like mode I opening and mode II shearing mode, most likely however as a mixture of both. We calculate the stress profiles with the components for mode I, the peeling stress and for mode II, the yz shear component along the edge (Path 1) and the center line (Path 2) on the adhesive bond-line in the z-direction (see Fig. 13 right) after the second de-moistening. The results show the effect of adhesive stiffness on the stress state (see Fig. 13 left).

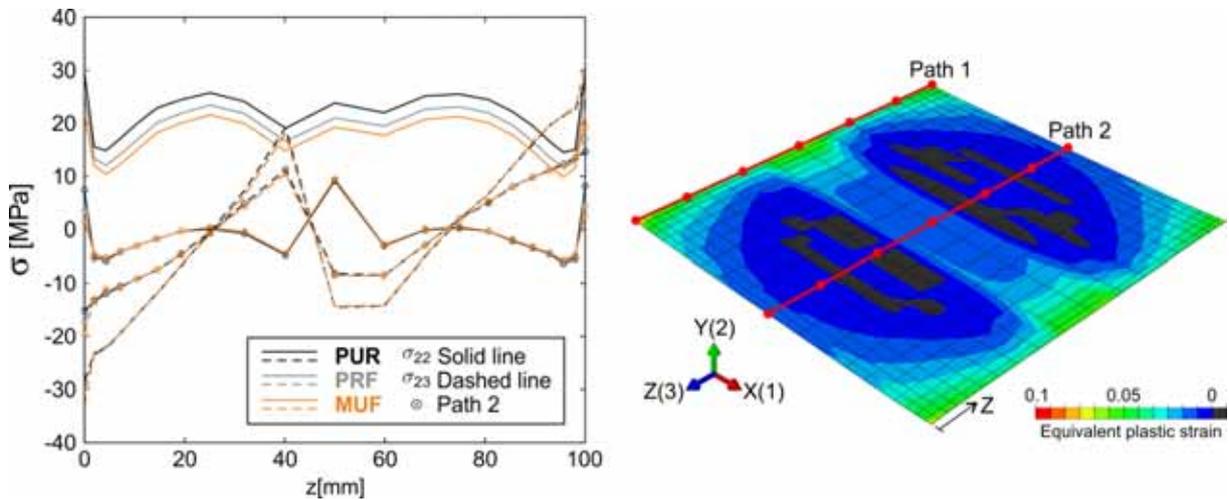

**Fig.** 13 Peeling and shear stresses in the adhesive bond-line along Path 1 and Path 2 for different types of adhesive (left) and the equivalent plastic strain in the lower PUR adhesive layer (right) of the P2-T3 sample after the second de-moistening

Delaminations are much more severe than transverse cracks in terms of loss of stiffness of the laminate, as they result in large stress-free regions and reduce the shear lag zones significantly. Additionally, at the delamination fronts, the loading on the adhesive bond-lines becomes maximal. Literature values for the critical energy release rates $G_{IC}$ and $G_{IIC}$ of adhesive bond-lines in beech wood unfortunately are given with a substantial bandwidth. We, therefore, pick a typical value used in Serrano (2004) and multiply both modes by a scalar multiplier.

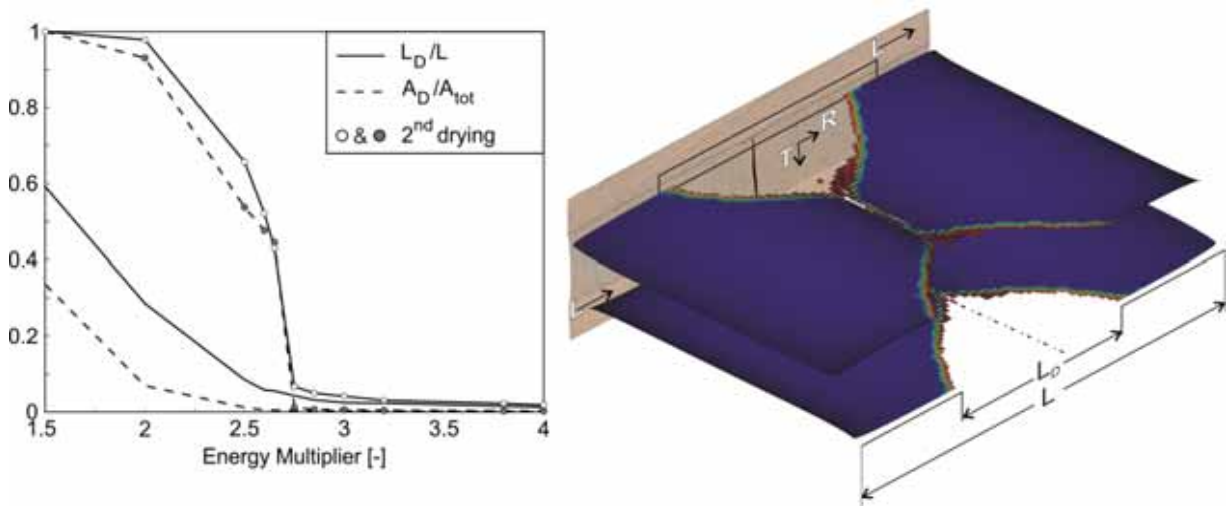

**Fig. 14** Relative delaminations in the adhesive bond-lines (left) and degradation scalar parameter (blue intact, red almost delaminated, while delaminated elements are deleted) with delaminations for a multiplier of 2.65 (right) compared to the P2-PUR-T3-3 sample

The calculated delamination ratio $(A_D/A_{tot})$ and edge delamination length $(L_D/L)$ for two successive cycles are shown as function of the multiplier in Fig. 14 left. It is evident that delamination evolves in consecutive cycles with a concave delamination front (see Fig. 14 right). Note that due to the curvature, pure fracture modes are not present, pointing at the importance of mixed mode criteria. In our case the best agreement between the edge delamination length and the corresponding experimental observation (see Fig. 10) is obtained by setting the multiplier between 2.65 and 2.75 with respect to values used in the linear elastic calculations (Serrano 2004), exhibiting the role of non-linear energy dissipation at the delamination front.

We observed damage evolution in the first two consecutive cycles experimentally, as well as in the calculations. Moisture gradients and swelling anisotropy both contribute to rather complicated stress states that relax by the buildup of the visco-elastic, mechano-sorptive or plastic strains. As a result, by inversion of the moisture state to the initial value, the sample does not return to the stress-free state, but can build up inverse stresses by the plastic deformation and mechano-sorption. We extrapolate in time from the experimentally studied 2 cycles up to 10 cycles computationally, corresponding to 3 years in practice. This is demonstrated in two typical scenarios that both start at the stress-free state, but case (I) is initially wet and then dries, while case (II) is initially dry and goes first through a moistening cycle. For clarity only stresses and strains in the center point (Point 2 in Fig. 11) in the radial direction with suppressed damage are visualized in Fig. 15.

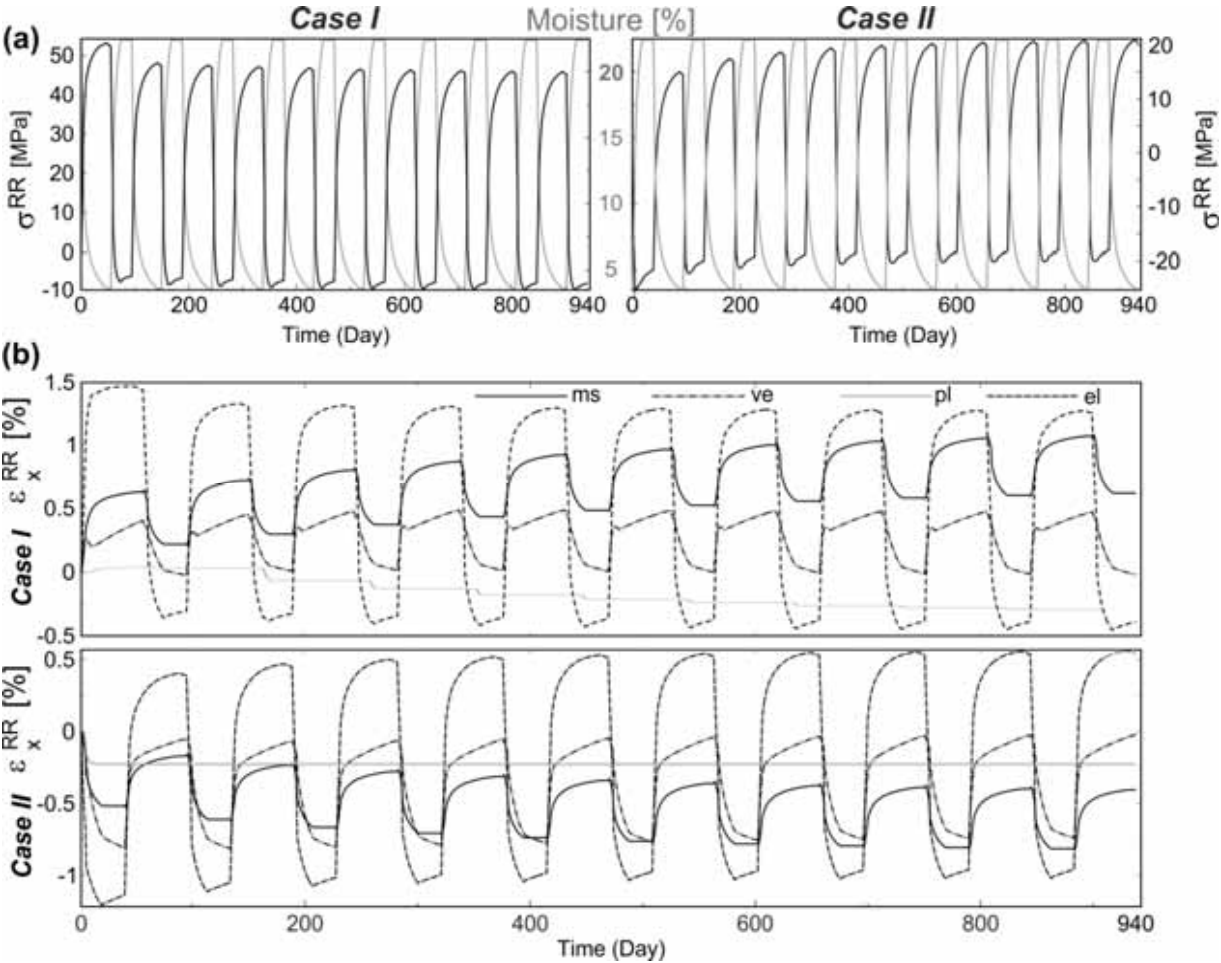

**Fig. 15** (a) Radial stress evolution in 10 consecutive drying-moistening (case I, left) and moistening-drying (case II, right) cycles at the center of the P2-T3 sample and (b) corresponding strain evolution in drying-moistening

(case I, top) and moistening-drying (case II, bottom) cycles (with identical legends). Subscript X represents ms, ve, pl, or el for the respective strain components

***Case I*** first dries, resulting in large tensile stresses $\sigma_{RR}$ that relax by the evolution of the visco-elastic and mechano-sorptive strains. When the sample is re-moistened the stress state inverts, leading to the compressive stresses and dissipation by plastic strains, as the moisture-dependent yield body shrinks due to moisture increase. In consecutive moisture cycles, the added tensile stress due to the plastic strain increment is over-compensated by the increased visco-elastic strain and more importantly by the mechano-sorptive one. During every re-moistening phase an additional but decreasing plastic strain increment is added. In total this causes a cyclic decrease of the maximum tensile stress that is becoming less significant after 5-6 cycles (see Fig. 15).

***Case II*** corresponds to adhesive bonding at rather dry states. When the sample is moistened for the first time, a significant plastic strain increment is formed while the central layer is in a compressive state. In re-drying, due to the expansion of yield surfaces, the inverted stresses cannot relax, resulting in significant tensile stresses. During consecutive cycles, the plastic strains remain constant, but mechano-sorptive and visco-elastic ones evolve, leading to cyclic tensile stress accumulation upon re-drying that does exceed the strength values. This can be considered as the driving mechanism behind hygro-fatigue (see Fig. 15).

## 3.3. Analytical prediction of micro-cracking in cross-laminated samples

Up to this point, we artificially introduced cracks in the middle lamella at experimentally observed positions and calculated the evolution of the strain fields as well as delamination initiation and propagation. In principle, one could extend the material model with capabilities for softening under tension or shear to localize micro-cracks. However, in this study we try to adopt a much more general fracture mechanical approach that is not limited by system sizes and is capable of predicting size effects.

The most suitable analytical solution to the damage evolution in cross-laminated samples is given by a micro-mechanics of damage approach, originally developed for transverse ply cracking and micro-delamination in cross-ply composites (Nairn 1989; Liu and Nairn 1990; Nairn and Hu 1992). Since this approach is considered as a standard in composite design for at least two decades, we only refer to the review articles in Nairn and Hu (1994) for a detailed description. It consists of two steps: (a) the use of micro-mechanics to analyze the stress field in a composite in the presence of damage determining the unit cell by its limiting cracks (see Fig. 10 inset) and (b) the use of a failure criterion to predict the evolution of damage. The equations for calculating the stress field in the two-dimensional (2D) unit cell are derived by variational mechanics. For high stiffness contrast of adjacent lamellae, like in the case of beech $(E_L / E_R)$, the sample edge can limit the unit cell.

The process is driven under drying in analogy to thermal cycling by differences in the hygro-expansion of the lamellae. Energy can be dissipated either by the formation of new micro-cracks, increasing the crack density, or by the initiation of delaminations at the tips of existing micro-cracks. This is expressed by the competition between the respective energy release rates (ERR) $G_{mc}$ and $G_{dc}$ for intra- and inter-laminar failure. Once $G_{dc}$ dominates $G_{mc}$, micro-delaminations form and the saturation crack density is reached only allowing for further energy dissipation by delamination growth. The model predicts the stress states in the middle as well as the ones in the outer lamellae, resulting in their cracking and delamination from the middle lamella under drying as well (Fig. 16). For simplification, we consider the material as transversely orthotropic, linear elastic with properties that correspond to the dry state.

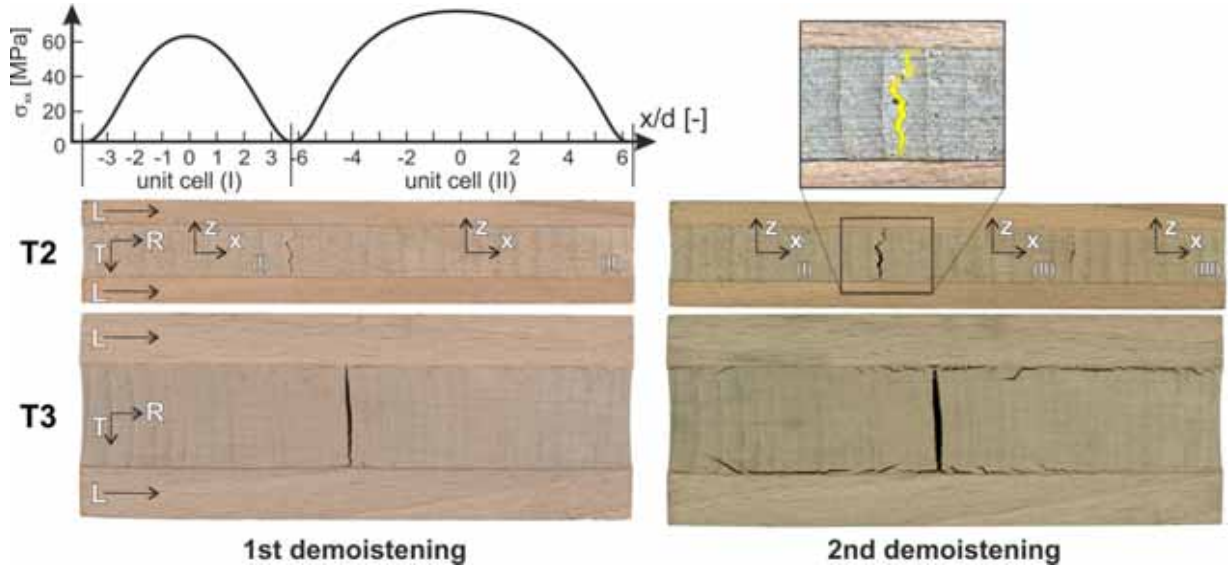

**Fig. 16** Side views of the fractured P2-PUR-T2-1/T3-3 samples after the 1st and 2nd de-moistening. The detail shows the microcrack-induced delaminations emanating from the micro-crack tips after the 2nd drying. $\sigma_{xx}$ is the average tensile stress along the x-direction in the middle lamella from the analytical solution (Nairn and Hu 1994)

A first micro-crack forms during the first drying in a region of more or less constant tensile stress $\sigma_{xx}$. As one can see in the stress profiles (see Fig. 16) the material assumptions lead to stresses above known strength values. However, verification calculation with the FEM approach that only considers the hygro-elastic part of our material model gives rise to similar stress values. Hence, the analytical approach will have mainly qualitative character. Nevertheless, the framework of micro-mechanics of damage helps in rationalizing the damage evolution. It is only during the second drying step that the strength of the middle lamella is overcome and segment II (in Fig. 16 T2) fractures, preferably in the mid span, in the region of maximum tensile stress as predicted by the model resulting in N=3 segments. Since also micro-delaminations form, $\rho^{10}$=(N/L)=(3/100)=0.03 mm$^{-1}$ is taken as the critical micro-crack density for this sample. Once $\rho^{10}$ is known, the critical intra-laminar ERR ($G_{mc}$) is determined and its relation with the inter-laminar one ($G_{dc}$) can be tuned such that $\rho^{10}$ emerges leading to $G_{dc}^{10} = 0.45 \cdot G_{mc}^{10}$ (see Fig. 17). As it can be seen, during the first de-moistening micro-cracking is the preferred damage mechanism. Since now all parameters are determined, the behavior of other thicknesses can be calculated (Fig. 17). The predicted evolution of the micro-crack density is given in Fig. 17, showing that for an identical moisture change of $\Delta\omega = 20\%$ the resulting crack density for d=20 mm is $\rho^{20}$=0.02 mm$^{-1}$ what is in agreement with the experiments (see Fig. 10). As a matter of fact, the accurate equilibrium critical inter-laminar ERR can be obtained as $G_{dc}^{20} = 0.71 \cdot G_{mc}^{10}$, what is close to $G_{dc}^{10}$ considering the finite size of the samples. Also it becomes evident that the threshold for micro-cracking is not reached for T1 samples. When we plot all critical moisture changes for the onset of micro-cracking as function of the thickness, we can extract the critical thickness for $\Delta\omega = 20\%$ to be about d=6 mm, presented in Fig. 17 inset as the limit line (bold dashed line).

Micro-mechanics of damage is a simple 2D approach and cannot capture the effect of the third direction. Considering the simplified material behavior and the limited dimensions of the samples, the predictability of the model and its quantitative agreement is surprisingly good. However, the obtained critical energy release rates are strongly dependent on the anatomic orientations, what should be considered in a study with broader experimental basis.

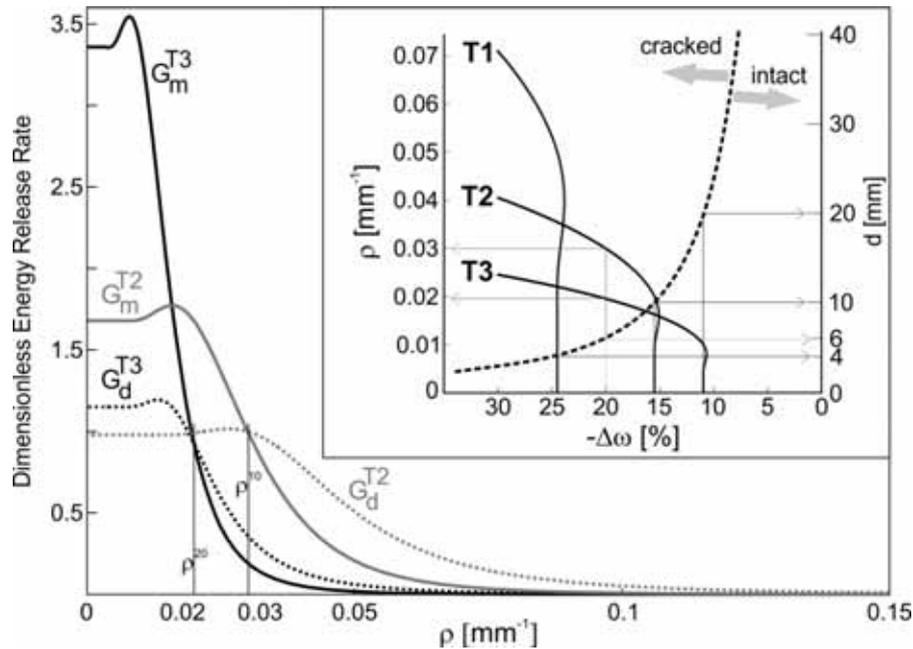

**Fig. 17** Dimensionless ERR vs. micro-crack density fitted to T2 and applied to T3 samples, giving excellent agreement with the experimental observations. Inset: Micro-crack density as function of the moisture gradient for P2-PUR-T1/T2/T3 samples and the limit line for micro-cracking initiation

## 4. Conclusions

The current work demonstrated that delaminations in hardwood bonding under changing climate can be studied on configurations and also under conditions that are much closer to real applications than typical delamination tests like the DIN EN 14080 for hardwood. In our study the aim was set on covering a large parameter space, what led to larger scatter of results by nature. Similar to the tensile-shear tests, the scatter can be drastically reduced by a careful selection of samples. Note that all adhesive bonds were produced in a moist condition, outside of the specifications of the respective adhesives, to obtain more severe crack and delamination states. However, calculations show (Fig. 15) that in principle also samples bonded at the dry states or lamellae with moisture differences at bonding can be covered to extend the reach of this study from a phenomenological one to a pre-study for alternative delamination testing of adhesive bond-lines in hardwood, that are closer to applications than today's strongly criticized codes.

Furthermore, it was indicated that due to high moisture gradients across the bond-line, resulting from the noticeable vapor resistance of the used adhesive systems, significant inelastic deformations and consequently local stresses build up under moisture cycling that lead to inter- and intra-laminar damage. By 3D calculations with an elaborated wood model we showed changes in the strain field in the region of the crack tips that are important for delamination initiation. Moreover, it was demonstrated how visco-elastic and mechano-sorptive strains relax upon crack formation, how plastic strains build up (see Fig. 12), and how the strain evolution in the moist states determines the stress situation upon drying, where crack initiation is observed in practice. The potential of stress dissipation within bond-lines to avoid delamination initiation was illustrated by calculating identical situations with different adhesive material models.

In addition, it was shown within a certain range of thickness d≤20 mm, how an adopted micro-mechanics of damage approach can be applied, for comparable growth ring orientations. This provides limit thickness values for the onset of micro-cracking and the values of adhesive bond-line delamination ratios under conditions that are similar to practical situations. This approach can be directly utilized by producers of CLT and veneer ply wood to calculate the range of application of specific laminates for climatic changes. It is our belief that the numerical model has a large potential for optimizing even hybrid CLT configurations with respect to the composition (lamella orientation and thickness) and process parameters such as moisture differences among lamellae.

# Appendix

**Table 2** Processing parameters of the three adhesive systems utilized to assemble the laminated panels

| Adhesive type | MUF | PRF | PUR |
|---|---|---|---|
| Adhesive name | Kauramin 683 + hardener 688 | Aerodux 185 RL + hardener HP155 | Purbond HBS 709 |
| Amount of adhesive application (g m$^{-2}$) | 340-440 (one side) | 225 (per side) | 180 (one side) |
| Mixing ratio | 100 g adhesive + 20 g hardener | 100 g adhesive + 20 g hardener | - |
| Pressing time (h) | 18.5 | 4 | 3 |

**Table 3** Coefficients for calculation of moisture-dependent Young's modulus for different adhesive systems fitted to the data published in Kläusler (2013). The value of the Poisson's ratio for all adhesive types is taken as $v_{adh} = 0.3$

| $E_{adh}(\omega) = a_0 + a_1\omega + a_2\omega^2 + a_3\omega^3$ | | | | |
|---|---|---|---|---|
| | $a_0$ (MPa) | $a_1$ (MPa/%mc) | $a_2$ (MPa/%mc$^2$) | $a_3$ (MPa/%mc$^3$) |
| **MUF** | 5355.0 | -604.100 | 33.270 | -0.6805 |
| **PRF** | 4176.0 | -176.90 | 19.38 | -0.8521 |
| **PUR** | 1242.0 | -158.300 | 26.250 | -6.443 |

**Table 4** Moisture expansion coefficients (CME) of different generic adhesive types given in Zhang et al. (2005)

| CME | $\beta$ (1/%mc) |
|---|---|
| MUF | 0.00198063 |
| PRF | 0.00172869 |
| PUR | 0.00171628 |

**Table 5** Normal entries of the visco-elastic compliance tensor pertaining to a serial association of six *Kelvin-Voigt* elements for different adhesive systems identified in Touati and Cederbaum (1997a) and Touati and Cederbaum (1997b)

| MUF | | | PUR | | |
|---|---|---|---|---|---|
| $i$ (-) | $J_i$ (MPa$^{-1}$) | $\tau_i$ (hour) | $i$ (-) | $J_i$ (MPa$^{-1}$) | $\tau_i$ (hour) |
| 1 | 1.94658E-06 | 1e-4 | 1 | 4.63471E-06 | 1e-4 |

| 2 | 0.755417E-06 | 1e-3 | 2 | 1.79861E-06 | 1e-3 |
|---|---|---|---|---|---|
| 3 | 1.13317E-06 | 1e-2 | 3 | 2.69802E-06 | 1e-2 |
| 4 | 1.40612E-06 | 1e-1 | 4 | 3.34791E-06 | 1e-1 |
| 5 | 2.59639E-06 | 1e0 | 5 | 6.18187E-06 | 1e0 |
| 6 | 1.00645E-06 | 1e1 | 6 | 2.39631E-06 | 1e1 |

**Table 6** Coefficients for calculation of moisture-dependent strength value $S_y(\omega)$ and non-linear hardening stress function $q(\omega)$ for PUR adhesive fitted to the data published in Kläusler (2013)

| $S_y(\omega) = b_0 + b_1\omega + b_2\omega^2 + b_3\omega^3$ | | | | |
|---|---|---|---|---|
| | $b_0$ (MPa) | $b_1$ (MPa/%mc) | $b_2$ (MPa/%mc$^2$) | $b_3$ (MPa/%mc$^3$) |
| | 29.21 | -7.734 | 4.696 | -0.9505 |
| $q(\omega) = (C_0 + C_1\omega)\alpha^{C_2}$ | | | | |
| | $C_0$ (MPa) | $C_1$ (MPa/%mc) | $C_2$ (-) | |
| | 60.3030 | -12.1352 | 0.5326 | |

**Table 7** Parameters for calculation of moisture-dependent diffusion coefficients for different adhesive systems following Volkmer et al. (2012) and Wimmer et al. (2013)

| $D(\omega) = D_0 e^{\alpha_0 \omega}$ | | |
|---|---|---|
| | $D_0$ (mm$^2$/hour) | $\alpha_0$ (-) |
| **MUF** | 9.792E-04 | 0.231 |
| **PRF** | 4.047E-04 | 0.231 |
| **PUR** | 3.067E-03 | 0.057 |

**Table 8** Parameters for the non-linear traction-separation behavior of cohesive element for PUR

| Damage initiation parameters (Serrano 2000) | | |
|---|---|---|
| $T_{nn}$ (MPa) | $T_{ss}$ (MPa) | $T_{tt}$ (MPa) |
| 6.55 | 14.2 | 14.2 |
| Damage evolution parameters / Fracture energies (Serrano 2004) | | |
| $G_I$ (N/mm) | $G_{II}$ (N/mm) | $G_{III}$ (N/mm) |
| 0.75 | 1.45 | 1.45 |
| Stiffness values | | |
| $K_{nn}$ (MPa) | $K_{ss}$ (MPa) | $K_{tt}$ (MPa) |
| 1157900 | 445400 | 445400 |

# Acknowledgement

This work was funded by the Swiss National Science Foundation in the National Research Programme NRP 66 - Resource Wood under grant No. 406640-140002: Reliable timber and innovative wood products for structures.